\documentclass[smallextended,natbib]{svjour3} 

\smartqed

\usepackage{amsmath,amssymb,amsfonts}
\usepackage{textcomp}
\usepackage{pifont}
\usepackage{fontawesome5}

\usepackage{graphicx}
\usepackage{xcolor}
\usepackage{tikz}
\usepackage{pgf-pie}
\usetikzlibrary{positioning, arrows.meta, fit, calc, backgrounds, shapes.arrows}

\usepackage{tabularx}
\usepackage{rotating}
\usepackage{booktabs}
\usepackage{makecell}

\usepackage{algorithm}
\usepackage{algpseudocode}
\usepackage{algorithmicx}
\usepackage{listings}
\usepackage[most]{tcolorbox}
\tcbuselibrary{listingsutf8}

\usepackage{url}
\usepackage{soul}

\usepackage{hyperref}
\definecolor{diffgreen}{RGB}{34,139,34}
\definecolor{diffred}{RGB}{178,34,34}
\definecolor{diffblue}{RGB}{25,25,112}
\definecolor{elegantbg}{RGB}{250,250,252}
\definecolor{bordercolor}{RGB}{100,100,100}

\lstdefinestyle{gitdiff}{
    basicstyle=\scriptsize\ttfamily,
    columns=fullflexible,
    keepspaces=true,
    breaklines=true,
    showstringspaces=false,
    frame=single,
    framerule=0.6pt,
    rulecolor=\color{bordercolor},
    backgroundcolor=\color{elegantbg},
    xleftmargin=0.8em,
    xrightmargin=0.8em,
    aboveskip=0.6em,
    belowskip=0.6em,
    moredelim=[is][\color{diffgreen}\bfseries]{+}{+},
    moredelim=[is][\color{diffred}\bfseries]{-}{-},
    moredelim=[is][\color{diffblue}\bfseries]{@@}{@@},
}

\definecolor{ghblue}{RGB}{130, 150, 190}
\definecolor{jp_purple}{RGB}{84, 9, 218}

\begin{document}

\newcommand{\tn}{ComBeautify}
\title{\tn{}: Leveraging Code Change Summaries for Conventional Commit Generation}
\titlerunning{\tn{}}

\author{
Rishi Raj, Aashray Tandon, Nikolay~Matyunin, Jibesh Patra
}

\institute{Rishi Raj, Aashray Tandon, Jibesh Patra\at
              IIT Kharagpur, India \\
              \email{rishiraj45035@gmail.com, aashraytandon1000@gmail.com, jibesh.patra@gmail.com}           %
           \and
           Nikolay Matyunin, \at
           Honda Research Institute Europe GmbH, Germany \\
              \email{nikolay.matyunin@honda-ri.de}
}

\date{}

\maketitle
\begin{abstract}
Modern software development extensively relies on Version Control Systems (VCS) such as Git to track changes and facilitate collaboration. While commit messages are crucial for conveying developer intent, they do not directly affect program functionality; as a result, developers often lack motivation to write meaningful commit messages.
The Conventional Commits Specification (CCS) addresses this issue by introducing a structured format and guidelines that help developers express the intent of a change, improving clarity, and enabling automation. Despite these specifications, developers frequently struggle to express intent while adhering to the format. In particular, they may misclassify commits due to the large number of possible categories or the lack of project-specific guidance. In this paper, we introduce \tn{}, an approach that automatically generates CCS-compliant commit messages. We use a hybrid approach that leverages context-preserving summarization of Git diffs with large language models (LLMs) to improve commit classification accuracy under CCS. Rather than relying on manually curated category definitions, we learn them from historical commits. We also propose a method for automatically generating the CCS description component.

Our extensive empirical evaluation across 13 languages, 6,500 open-source repositories, and over 20 million commits demonstrates long-term insights into CCS adoption trends, revealing increasing compliance.
Our evaluation on twenty popular repositories reveals the limitations of approaches guided by generic category definitions, and shows that \tn{}, when trained on project-specific data, outperforms baseline approaches by up to 3$\times$ in some cases.
Additionally, using a dataset of 2{,}000 manually curated commits, we validate our summarization approach, achieving up to \textbf{95.25\%} accuracy and surpassing baseline methods.
These findings emphasize the need for adaptable, context-aware AI methods for commit classification and generation of CCS-compliant commit messages.
\end{abstract}

\section{Introduction}\label{sec:intro}
Version Control Systems (VCS) such as Git, Subversion, and Mercurial have become an integral part of the modern software development workflow. They allow multiple developers to collaborate on a project, keeping track of the modifications to the codebase. A commit is a fundamental operation that captures a snapshot of changes made to the codebase at a particular point. In addition to changes in the codebase, a commit contains additional information such as the author of the change, the time and date of the commit, and a commit message typically written in a natural language. The commit message is primarily used to express the developer's intent behind the change and to communicate it with other developers.

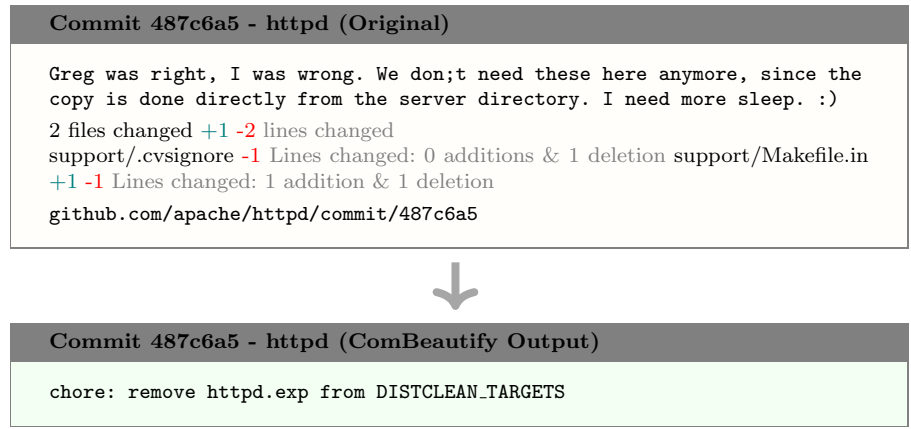
\begin{figure}[ht]
  \centering
  \begin{tcolorbox}[
      title=\textbf{Commit 487c6a5 - httpd (Original)},
      colback=yellow!2!white,
      colframe=black!50,
      coltitle=black,
      fonttitle=\bfseries,
      fontupper=\small,
      enhanced,
      boxrule=0.5pt,
      sharp corners
    ]

    \texttt{Greg was right, I was wrong.
      We don;t need these here anymore, since the copy is done directly from the
      server directory.
    I need more sleep. :)
  }

  \vspace{0.3em}
  2 files changed \textcolor{teal}{+1} \textcolor{red}{-2} \textcolor{gray}{lines changed}

  support/.cvsignore \textcolor{red}{-1} \textcolor{gray}{Lines changed: 0 additions \& 1 deletion}
  support/Makefile.in \textcolor{teal}{+1} \textcolor{red}{-1} \textcolor{gray}{Lines changed: 1 addition \& 1 deletion}

  \vspace{0.3em}
  \texttt{github.com/apache/httpd/commit/487c6a5}
\end{tcolorbox}

\vspace{-3pt}

\begin{tikzpicture}
  \draw[->, draw=gray!70, line width=3pt, shorten >=1pt, shorten <=1pt] (0,0.3cm) -- (0,-0.4cm);
\end{tikzpicture}
\vspace{-5pt}
\begin{tcolorbox}[
    title=\textbf{Commit 487c6a5 - httpd (\tn{} Output)},
    colback=green!5!white, %
    colframe=black!50,
    coltitle=black,
    fonttitle=\bfseries,
    fontupper=\small,
    enhanced,
    boxrule=0.5pt,
    sharp corners
  ]

  \texttt{chore: remove httpd.exp from DISTCLEAN\_TARGETS}
\end{tcolorbox}

\caption{Commit message from the \textit{httpd} repository (top), and its transformation into a CCS-compliant message (bottom) by our approach, \tn{}. The original message is informal and lacks technical detail, while the generated message adheres to CCS with a clear type and description.}
\label{fig:commit_combined}
\end{figure}

Since commit messages themselves do not directly impact any functionality brought by the changes, developers often lack the motivation to write a proper commit message. Consider the commit \texttt{487c6a5} from the \textit{httpd} repository of the \textit{apache} project shown in Figure~\ref{fig:commit_combined} (top).
This message is informal, self-reflective, and lacks a clear technical explanation of the changes made. It references internal team dynamics and personal circumstances rather than providing an objective summary of the modification and as a result such messages are often discouraged.

To this end, ~\citet{tian2022makes} propose a criterion for a high-quality commit message, emphasizing that it should consist of both `what' (summary of changes in the commit) and `why' (reasons for making those changes). Their analysis of five open source projects finds that 44\% of the commit messages lack essential information, of these 44\% contain only `what', 12\% only contain `why', and 4\% contain neither `why' nor `what'. This highlights a prevalent issue in modern software development: writing a highly effective commit message. This can be due to developers deprioritizing writing high-quality messages due to time constraints and the belief that changes are self-explanatory. In addition to this, there is a lack of clear guidelines on what constitutes a good commit message in many projects, and distributed teams with varying practices further contribute to this inconsistency. These challenges highlight the need for an automated tool to help developers write
highly effective commit messages.

The lack of a standardized commit format often leads to fragmented and ambiguous commit messages, hindering effective repository management and code understanding. To address this, the Conventional Commits Specification (CCS) introduces a standardized, project-agnostic format. It specifies that a commit message should be structured as \texttt{<type>[optional scope]: <description>}, followed by an optional body and footer. For example, consider the commit \href{https://github.com/angular/angular/commit/3839cfbb18fcc70cae5a6ba4ba7676b1c4acf7a0}{\texttt{3839cfb}} taken from the GitHub repository of Angular shown in Figure~\ref{fig:commit_following_CCS} (and abbreviated here for brevity).
The \texttt{<type>} in this case is \textit{fix} and the \texttt{<description>} is \textit{``Routed components never inherit RouterOutlet EnvironmentInjector"}. The commit fixes a bug in response to an issue report. The intention of the developer is made clear through the use of the \textit{fix} type and a one-line description to explain the change. \textit{router} represents the \texttt{[scope]} which provides additional context to the type. The remaining part of the commit message contains the body and the last line represents the footer.

\begin{figure}[ht]
\centering
\begin{tcolorbox}[
    title=\textbf{Commit 3839cfb}, %
    colback=gray!5!white,
    colframe=black!50,
    coltitle=black,
    fonttitle=\bfseries,
    fontupper=\small,
    enhanced,
    boxrule=0.5pt,
    sharp corners,
  ]

  \textbf{fix(router)}: Routed components never inherit \texttt{RouterOutlet} EnvironmentInjector

  \vspace{1em}

  This commit ensures components in the route config predictably
  always get their providers from the hierarchy available to routes rather than sometimes being dependent on where they are inserted.

  \vspace{0.5em}
  fixes: \href{https://github.com/angular/angular/issues/53369}{\#53369}

  \vspace{0.5em}
  BREAKING CHANGE: Providers available to the routed components always come from the injector hierarchy of \ldots

  \vspace{0.5em}
  PR Closed: \href{https://github.com/angular/angular/pull/54265}{\#54265}

\end{tcolorbox}
\caption{Sample Conventional Commit Message.} %
\label{fig:commit_following_CCS}
\end{figure}

The previous example (Figure \ref{fig:commit_following_CCS}) demonstrates some benefits of following CCS. First, a developer reading the commit message can immediately infer (from the \textit{type}) that the commit addresses a specific issue, and receive a concise indication of the scope and nature of the fix. Second, the structured format of the message enables automated tools to parse and utilize the information effectively for a variety of purposes.
We analyze selected usages of CCS in Section~\ref{subsec:case_studies_ccs}.

To this end, this paper aims to automatically synthesize a structured conventional commit message from a non-compliant commit.
We propose an approach that employs Large Language Models (LLMs) to summarize commit changes and combines this summary with the original commit message to generate a complete CCS-compliant commit. Specifically, we use two fine-tuned LLMs to generate the CCS type and description, respectively.
As a result, our approach produces an informative, structured commit message that complies with the CCS specification.
The bottom part of Figure~\ref{fig:commit_combined} shows an example of a CCS-compliant output generated by our approach.

A key insight of our approach is that in practice CCS conventions are adapted to project-level specifics.
Whereas prior work on commit classification typically targets a predefined, manually curated set of commit  types~\citep{zeng2024first}, our approach automatically learns each project's own conventions from its commit history, and uses them to synthesize both the type and the description of a compliant message. We demonstrate that such project-specific learning outperforms generalized models for commit classification and commit message generation.

In summary, the primary contributions of this paper are as follows:

\begin{itemize}
\item[(1)] \textit{Adoption:} We present a large-scale analysis of the current landscape of the CCS adoption by developers and which languages and developer
  communities are primarily adopting it.
\item[(2)] \textit{CCS-compliant commits:} We present an approach to generate CCS-compli\-ant commit messages from non-conforming commits, aligned with project conventions,. 
\item[(3)] \textit{Case study:} We present a case study on how developers leverage the available structure of a commit for a variety of purposes, such as changelog generation and semantic version bump.
\end{itemize}

\section{Approach}
\label{sec:approach}
This section introduces~\tn{}, our approach to the generation of a structured commit message compliant to the Conventional Commit Specification (CCS).

\subsection{Overview}\label{subsec:problem_statement}
The goal of \tn{} is to synthesize a conventional commit message from a non-conforming commit.
We adopt a two-step approach for~\tn{}.
First, instead of directly incorporating the code changes, we create a summary and combine it with the original commit message. In the next step, we use this combination to generate the required type and description complying with CCS.
Specifically, we define elements of a commit and its conventional version as follows:

\begin{definition}
  \textbf{Commit}: A commit $C$ is a tuple $(m, F)$  where the $m$ is the commit message and $F = \{f_1, f_2, \ldots f_n\}$ are the changed files involved in the commit $C$.
\end{definition}

\begin{definition}
  \textbf{Conventional Commit}: A conventional commit $C_c$ is a tuple $([t, d], \mathbb{B}, \iota, F)$  where  $F = \{f_1, f_2, \ldots f_n\}$ are the changed files involved in the commit, $t$ is the required \textit{type} and $d$ is the required \textit{description} as mandated by CCS. $\mathbb{B}$ and $\iota$ are the optional body and footer components of $C_c$.
\end{definition}

The input to \tn{} is an non-compliant commit $C$, and the goal of \tn{} is to automatically generate a commit $C_c$ from $C$.
We focus on two major challenges faced by developers in adhering to the structured specification of conventional commits: assigning the type and writing the description~\citep{zeng2024first}.
Since the body ($\mathbb{B}$) and the footer ($\iota$) are optional for $C_c$, their generation is considered out of scope for this work.

Algorithm ~\ref{alg:conversion} outlines the main steps of the proposed approach. In the addition to the commit $C$ as input, the algorithm utilizes three additional components: $\text{LLM}_{summary}$, an LLM-based  summarizer of changes introduced in the commit; $\text{LLM}_{type}$, an LLM-based type generator; $\text{LLM}_{description}$, an LLM-based description generator.
For each file involved in a commit, the algorithm first computes the file  differences
using the DIFF function, and aggregates them into  $F_{\Delta}$ for all files in the commit $C$ (lines 2--4).
Commit differences can be arbitrarily large, which  posses a challenge for effective comprehension. As a result, instead of using the diff directly, we compute a natural language summary $S$ of the differences using an LLM (line 5).  The obtained summary is then combined with the original commit message (line 6), and provided as input to two separate LLMs to obtain the type (line 7) and the description (line 8). Finally, we return the synthesized structured commit $C_c$.
The rest of the section provides details on each step of our approach.

\begin{algorithm}[t]
  \caption{Convert an unstructured commit to CCS}
  \begin{algorithmic}[1]
    \Require An unstructured Commit message $C=(m, F)$, $\text{LLM}_{summary}$, $\text{LLM}_{type}$, $\text{LLM}_{description}$
    \Ensure Structured Commit $C_c$ that complies to CCS
    \State $F_{\Delta} \leftarrow \{\}$
    \For{$f_i$ in $F$}
    \State $F_{\Delta} \leftarrow \text{DIFF}(f_i)$ \Comment{Aggregate file-level differences}
    \EndFor
    \State $S = \text{LLM}_{summary}(F_{\Delta})$ \Comment{Get summary of the diff}
    \State $\mathbb{C} = m + S$ \Comment{Append summary to the message}
    \State $t = \text{LLM}_{type}(\mathbb{C})$ \Comment{Get type}
    \State $d = \text{LLM}_{description}(\mathbb{C})$ \Comment{Get description}
    \State $C_c = ([t,d], F) $
    \State \Return $C_c$
  \end{algorithmic}
  \label{alg:conversion}
\end{algorithm}

\subsection{Creating Commit Summary}
The primary elements of a commit consist of one or more file changes along with a message that describes the nature of the change. These file changes are typically represented as file-based differences (diffs), which capture modifications at the code level. To this end, effectively capturing the developer's intent behind a commit presents two key challenges:

\begin{itemize}
  \item[(1)] File-level diffs of commits can often be arbitrarily large (the average character length of a diff in our dataset is 20,324), making them difficult to process comprehensively.
  \item[(2)] The mandatory components of a CCS, namely, the type and description, are generally expressed in natural language (e.g., English). Therefore, relying solely on file-based diffs, which emphasize code-level changes, is suboptimal for tasks involving natural language generation.
\end{itemize}

To address these challenges, our key insight is to generate a natural language summary of the diff and use this summary as input instead of directly providing code-level diffs.

\paragraph{File-level differences and summary.}
The example commit contained in Figure~\ref{fig:commit_combined} modifies two files \textit{support/.cvsignore} and \textit{support/Makefile.in}. The green digits prefixed with a \textcolor{teal}{$+$} indicate the number of lines added, while the red digits prefixed with a \textcolor{red}{$-$} indicate the number of lines deleted. By default, \texttt{git diff} uses the \textit{Myers} basic greedy diff algorithm to compute differences. As an initial step in generating the summary, we employ this default method to extract file-level differences.
 
To be effective, the summary must capture the key aspects of the code changes in a diff. In recent years, large language models (LLMs) have demonstrated remarkable code-understanding capabilities \citep{feng2020codebert,nam2024using,anand2024critical,chen2021evaluating}. Accordingly, we adopt an in-context learning approach to generate natural language summaries of code diffs. Specifically, the computed diff is tokenized and, together with a carefully constructed prompt and a few examples, is provided as input to $\text{LLM}_{summary}$, which generates a brief summary $S$ (in our case, the average summary length is 688 characters, a 30x reduction from the original code diff).

An example diff (shortened for brevity), the prompt used, and the LLM-generated summary are shown in Figure~\ref{fig:boxes}.
The prompt is crafted such that the LLM faithfully summarizes the diff. Additionally, we provide the LLM with two example diffs along with manually crafted summaries. This in-context learning helps the model effectively generate suitable summaries. As a result, we obtain
a summary that captures the key file changes and describes them structurally in natural language, abstracting away from the raw diff.

\begin{figure*}[htbp]
  \centering
  \begin{minipage}[t]{0.45\textwidth}
    \raggedright
    {\textbf{Input: Git Diff}}
\begin{lstlisting}[basicstyle=\ttfamily\scriptsize, frame=single]
diff --git a/ibis/tests/sql/
test_sqlalchemy.py
@@ -841,3 +841,63 @@
+@pytest.fixture
+def person():
+    return ibis.table(
+        dict(id="string",
+             personal="string",
+             family="string"),
+        name="person",
+    )
+
+@pytest.fixture
+def visited():
+    return ibis.table(
+        dict(id="int32",
+             site="string",
+             dated="string"),
+        name="visited",
+    )
+
+@pytest.fixture
+def survey():
+    return ibis.table(
+        dict(taken="int32",
+             person="string",
+             quant="string",
+             reading="float32"),
+        name="survey",
+    )
+ ...
\end{lstlisting}

  \end{minipage}
  \hfill
  \begin{minipage}[t]{0.50\textwidth}
    \small
    \textbf{Prompt}\\
    \textit{You are a code summarizer based on git diff. Given the git diff, please summarize it while keeping the following points in consideration:}
    \vspace{0.3em}
    \textit{
      \begin{enumerate}
        \item We will be using this summarized git diff to classify commits, so do not omit any important context.
        \item Avoid redundant information in the summary, as we want to minimize the number of output tokens but at the same time don't omit any useful context.
      \end{enumerate}}
    \vspace{0.3em}
    \textit{
      Output should be in this format:\\
      File1: \textless summarized points\textgreater\\
      File2: \textless summarized points\textgreater\\
      Here are few examples:\\
      \textless few shot examples \textgreater
    }
    \vspace{0.3cm}
    \hrule
    \vspace{0.3cm}
    \textbf{Output: Generated Summary}\\
    File: \texttt{ibis/tests/sql/test\_sqlalchemy.py}
    \begin{itemize}
        \setlength{\itemsep}{0.1em}
      \item Added three pytest fixtures defining table schemas: \texttt{person}, \texttt{visited}, and \texttt{survey}
      \item Implemented test function \texttt{test\_no\_cross\_join} to validate SQL join operations between the defined tables
      \item Test verifies proper join logic without generating unintended
        spacecross
        joins
    \end{itemize}
  \end{minipage}
  \caption{Example diff, prompt and the summary produced by $\text{LLM}_{summary}$.}
  \label{fig:boxes}
\end{figure*}

\paragraph{Combined commit message.}
While a diff captures code-level changes, the commit message typically conveys the developer's intent in natural language. Although there is no enforced limit on commit-message length, in practice these messages are quite short (the average length in our dataset is 91 characters). Therefore, we perform no additional processing on the message and simply append it to the diff summary obtained earlier. We refer to this combined representation as the \textit{commit-intent representation} ($\mathbb{C}$ in Algorithm~\ref{alg:conversion}).

The \textit{commit-intent-representation} $\mathbb{C}$ serves as the primary input for generating the mandatory components of a CCS-compliant message: the \textit{type} and \textit{description}.
We fine-tune two separate LLMs, one for type generation and the other for description generation. We decouple these tasks because they require different levels of abstraction. Generating a single-word type typically
requires understanding the developer's high-level intent. In contrast, generating a multi-word description demands a fine-grained understanding of the semantic impact of the change. For both fine-tuning approaches, we use commits that comply with CCS as input. We describe each task in more detail below.

\subsection{Generating Type}
The Conventional Commit Specification defines a type as a mandatory element of the commit message. A type, a single word according to the specification, signifies the kind of change made in the commit. The type can be optionally accompanied by a scope, which provides additional context. The focus of our work is to generate only the mandatory components of a CCS, and as a result, generating the scope is not considered in this paper.

The specification recommends, but does not strictly enforce, any type except two: \textit{feat} and \textit{fix}. The \textit{feat} type is used when a commit adds a new feature, while \textit{fix} is used for bug-fixing commits. The remaining types do not have formally specified definitions. Additionally, developers are free to introduce new custom types beyond the standard CCS types. The reasons for project-specific definitions and custom types are manifold. First, some projects have domain-specific nuances that are not captured by standard categories. Second, projects may rely on specific automated workflows based on internally agreed-upon types.

The lack of clear definitions of CCS types and the ability to create custom types result in confusion~\citep{zeng2024first} among developers on the correct type to use for a commit\footnote{\url{https://github.com/conventional-commits/conventionalcommits.org/issues/283}} and can potentially create friction
for new contributors unfamiliar with a project's conventions. Furthermore, project-specific types are often either undocumented or buried in contribution guidelines or README files.

This challenge is evident in how common development activities are categorized. To illustrate this, we analyzed our dataset for commits related to two frequent tasks: resolving merge conflicts and updating dependencies. For each task, commits were identified by searching commit messages for predefined phrases (e.g., ``fix merge conflict" and ``resolve conflicts" for merge conflicts; ``update dependencies" and ``bump deps" for dependency updates).
Table~\ref{tab:type-distribution-common-tasks} presents the distribution of CCS types assigned to commits performing these tasks, focusing on five popular  types: \texttt{feat}, \texttt{fix}, \texttt{chore}, \texttt{build}, and \texttt{refactor}.
Results demonstrate that  common development activities are inconsistently categorized across various CCS types. Such dispersion for well-defined tasks indicates ambiguity in CCS type selection and highlights  the need for automated assistance.

\begin{table}[htbp]
  \centering
  \caption{Distribution of CCS Types for Common Development Tasks. Counts indicate the number of commits found in our dataset for each task, for five popular CCS types.}
  \label{tab:type-distribution-common-tasks}
  \begin{tabular}{@{}lrrrrr@{}}
    \toprule
    Task & \multicolumn{1}{c}{feat} & \multicolumn{1}{c}{fix} & \multicolumn{1}{c}{chore} & \multicolumn{1}{c}{build} & \multicolumn{1}{c}{refactor} \\
    \midrule
    Merge Conflicts    & 546   & 413    & 428     & 30        & 92        \\
    Update Dependencies& 731   & 17462  & 46850   & 5072      & 131       \\
    \bottomrule
  \end{tabular}
\end{table}

To help mitigate these challenges for developers, we propose an approach to automatically generate the type. One of the major challenges for an approach that generates a type is the widespread use of custom types (there exist more than 215K distinct types in our dataset, Section~\ref{subsec:adoption_trend_ccs}). To address this challenge, a potential solution is to manually curate definitions for each type, as explored in prior work~\citep{zeng2024first} and use these definitions to guide a type-generating system. However, a manual strategy faces significant limitations:
\begin{itemize}
  \item[(1)] It does not scale to the vast and continuously growing number of different possible types.
  \item[(2)] The semantics of the same type can vary across projects which limits the suitability of a universal definition.
\end{itemize}

To overcome this limitation, we propose to learn project-specific CCS types from historical commits. Our key insight is that while the semantics of a given type tend to remain consistent within a single project, they can vary significantly across different projects. This variability posses a challenge for a type-generating system that relies on manually curated, predefined type definitions.  Additionally, a project-specific approach is not limited to a custom set of predefined types.

We obtain $\text{LLM}_{type}$, by fine-tuning an LLM using the \textit{commit-intent-repre\-sentation} (Section~\ref{sec:convert_non-ccs_to_ccs}) of historical commits. Specifically, $\text{LLM}_{type}$ is trained using $\mathbb{C}$ as input where the task is to generate $t$. For this step, we compute the commit-intent-representation $\mathbb{C}$ from a given conventional commit $C_c$ as follows:
\[
  \mathbb{C} =
  \left\{
    \begin{array}{l}
      d + \mathbb{B} + S \mid  \mathbb{B} \in C_c \\
      d +  S \mid  \mathbb{B} \notin C_c
    \end{array}
    \right.
  \]
  The symbol $+$ denotes simple string concatenation, and $S$ refers to the diff summary as described in Algorithm~\ref{alg:conversion}.

  \subsection{Generating Description}
  The description is another mandatory component of CCS. It typically consists of a short phrase that succinctly summarizes the nature of the change. To generate the description, we employ a dedicated LLM and fine-tune it specifically for this task. The model takes the commit-intent-representation as input and produces the corresponding concise description.

  One precondition for  choosing commits for fine-tuning $\text{LLM}_{description}$ is that they must contain a body $\mathbb{B}$ which is optional according to CCS. The body is typically written in natural language and contains a detailed explanation of the changes performed in the commit.

  $\text{LLM}_{description}$ is derived by fine-tuning an LLM using $\mathbb{C}$ as input where the task is to generate $d$. For this step, we compute the commit-intent-representation $\mathbb{C}$ from a given conventional commit $C_c$ as follows:
  $$\mathbb{C} = S + \mathbb{B},$$
  \noindent where $S$ is the diff summary as described in Algorithm~\ref{alg:conversion}.

  \subsection{Inference}
  Finally, we utilize $\text{LLM}_{type}$ and $\text{LLM}_{description}$  to generate the conventional commit $C_c$, as outlined in Algorithm~\ref{alg:conversion}.
  Both LLMs take the commit-intent-representation $\mathbb{C}$ of
  the commit as their input. This representation is subsequently fed into the fine-tuned $\text{LLM}_{type}$ to predict the commit type $t$
  and into $\text{LLM}_{description}$ to generate the description $d$.
  Finally, these two outputs, $t$ and $d$, are concatenated, along with the associated file changes $F$,
  to form the conventional commit
  $C_c$.

\section{Experimental Setup}\label{sec:setup}
\subsection{Benchmark}
We construct a dataset of 6,500 open-source repositories spanning 13 programming languages. For each language, we select the top 500 repositories by GitHub star~\ding{73} count. For repositories with multiple languages, we use GitHub's primary language label to assign a single language to each repository.

We develop an automated pipeline to collect and process commit-level data from these repositories, focusing on activity from January 2017 onward. The pipeline initially attempts shallow cloning to minimize storage and network overhead (\texttt{git clone --shallow-since}). If this fails, we use fallback mechanisms, including full cloning and GitHub’s REST API, with rotating tokens to mitigate rate limits.
As a result, the pipeline yields approximately 21.5 million non-merge commits from 6,500 repositories, constituting our experimental dataset.
Within this dataset, we identify CCS-compliant commits using regular expressions to detect messages that adhere to the \texttt{<type>: <description>} structure, optionally followed by a body.

We ran our experiments on a Rocky Linux machine equipped with an Intel Xeon E5-2650 v4 CPU @ 2.20\,GHz, NVIDIA L40 GPU and 64\,GB RAM. All experiments were implemented in Python 3.10.8 using the Hugging Face Transformers library. We performed supervised fine-tuning (SFT) with LoRA (PEFT) on a 4-bit quantized base model (bitsandbytes), using the Hugging Face Trainer.

\subsection{Pre-trained LLMs within \tn{}}

We use Code Llama Instruct 7B as the summarizer LLM ($\text{LLM}_{summary}$) to obtain context-aware git diff summaries. For deriving $\text{LLM}_{type}$, we use the base Code Llama 7B and fine-tune it for our purposes. Similarly, the description generating $\text{LLM}_{description}$ is fine-tuned based on Code Llama Instruct 7B.

\subsection{Empirical Study}\label{subsec:empirical_study}
\label{subsec:adoption_trend_ccs}
To assess the prevalence of conventional commits among developers, we conduct a large-scale empirical analysis of adoption trends across open-source projects.

\paragraph{Commit-level adoption.}
Our analysis reveals that, out of 20.76 million commits in our dataset, about 5.1 million are CCS commits, accounting for 24.58\%.
of the total, authored by over 600,000 unique contributors.

\paragraph{Repository-level adoption.}
Of the 6,500 repositories present in our dataset, we assess CCS adoption by calculating repository-level compliance as the percentage of  commits adhering to the Conventional Commits specification. Our analysis reveals that 1247 repositories (19.2\%) exhibit at least 70\% compliance, with 991 (15.2\%) surpassing 80\%, 708 (10.9\%) exceeding 90\%, and 482 (7.4\%) demonstrating compliance above 95\%.

\paragraph{Year-wise adoption.}
We also examine year-wise CCS adoption from 2017 onward across all commits in our dataset. In 2017, 16.3\% of commits were CCS-compliant. By 2025, this proportion had increased to 35.9\%, representing more than a twofold rise over eight years. This trajectory indicates a consistent increase in adherence to the CCS specification over time.

\begin{figure}[t]
  \centering
  \includegraphics[width=0.85\linewidth]{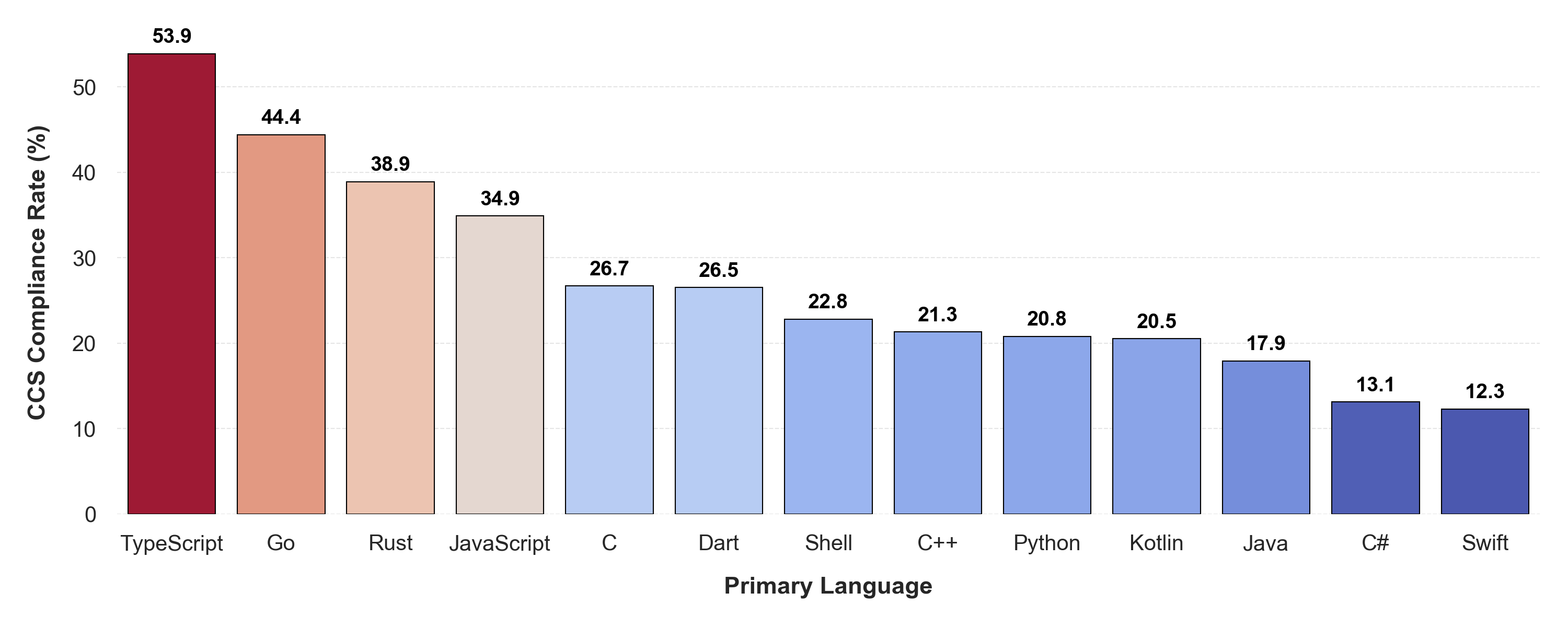}
  \caption{Average CCS compliance by programming language. Values represent the mean percentage of CCS-compliant commits per language.}
  \label{fig:language_heatmap}
\end{figure}

\paragraph{Adoption by Language.}
We further analyze CCS adoption across programming languages to assess whether certain ecosystems exhibit stronger adherence. Figure~\ref{fig:language_heatmap} summarizes the results. TypeScript projects show the highest average CCS compliance (53.9\%), followed by Go (44.4\%), Rust (38.9\%), and JavaScript (34.9\%). Dart, C, C++, Python, Shell, and Kotlin exhibit moderate adoption, whereas Java, C\#, and Swift show comparatively lower compliance rates.

\paragraph{Distribution of CCS Types.}
To understand the current landscape of the varied types that developers adopt during a commit, we analyze all CCS commits and check their types. The ten most frequent CCS types are \textit{fix (13.4\%), chore (10.3\%), feat (8.3\%), docs (4.2\%), refactor (2.7\%), build (2.6\%), test (1.7\%), ci (1.3\%), doc (1.1\%), tests (0.8\%)}
We find that CCS types form a long-tailed distribution. In total, there are over 215,000  distinct types. We find that due to the lack of restrictions by the specification there is an extensive use of custom CCS types.

\paragraph*{}
Overall, our empirical study confirms rising CCS adoption across languages and repositories and highlights diversity of CCS types used in individual projects. Together, these findings motivate the need for an automated approach to aid developers in formulating project-aligned CCS types and commit messages.

\section{Evaluation}
\label{sec:evaluation}

We evaluate~\tn{} on the following research questions:
\begin{itemize}
  \item \textit{RQ1:} What is the effectiveness of our approach in generating the type part of a conventional commit?
  \item \textit{RQ2:} What is the effectiveness of our approach in generating the description part of a conventional commit?
  \item \textit{RQ3:} How do projects take advantage of structured conventional commits?
  \item \textit{RQ4:} Is including an additional code summarization component in \tn{} useful?
\end{itemize}

\subsection{RQ1: Effectiveness of \tn{}
in Generating Type}\label{sec:convert_non-ccs_to_ccs}
The empirical study of  Section~\ref{subsec:empirical_study} show a steady increase in CCS adoption by developers. The findings indicate that although about 19\% of repositories have at least 70\% CCS compliance, there are still a large number of commits that do not comply with CCS.
To this end, we investigate whether~\tn{} can automatically transform unstructured commits into fully CCS-compliant commits.

\begin{table*}[!ht]
  \centering
  \caption{Performance of \tn{} in generating type and comparison to baselines.}
  \label{tab:model_performance_across_projects}
  \footnotesize
  \setlength{\tabcolsep}{2pt}
  \renewcommand{\arraystretch}{1.3}
  \resizebox{\linewidth}{!}{%
    \begin{tabular}{|l|c|c|c|c|c|c|c|c|>{\raggedright\arraybackslash}p{6.65cm}|}
      \toprule
      \makecell{Project} & \makecell{\faGithub\\stars} & \makecell{Guid-\\elines} & \makecell{NN-\\Gen} & \makecell{Copilot} & \makecell{Gemi-\\ni} & \makecell{$\text{manual}_{\text{def}}$\\ (\%)} & \makecell{$\text{global}_{\text{def}}$\\ (\%)} & \makecell{\tn{}\\ (\%)} & Top seven CCS types (frequency in \%) \\
      \midrule
      freeCodeCamp & 438K\ding{73} & \ding{52} & 64.75 & 44.28 & 43.50 & 72.75 & 24.00 & 76.00 & \textit{fix(47.79), chore(31.43), feat(13.79), docs(2.93), refactor(2.36), test(1.36), feature(0.36)}  \\
      deno & 106K\ding{73} & \ding{54} & 44.00 & 54.41 & 55.75 & 71.50 & 50.75 & 70.75 & \textit{fix(38.00), chore(17.79), feat(15.71), refactor(11.86), perf(4.43), docs(4.21), test(2.29)}  \\
      angular & 100K\ding{73} & \ding{52} & 49.75 & 53.32 & 50.00 & 71.25 & 58.00 & 70.75 & \textit{docs(22.79), fix(21.21), refactor(18.29), build(15.43), feat(8.29), test(5.36), ci(3.79)}  \\
      ant-design & 98K\ding{73} & \ding{54} & 45.00 & 53.38 & 45.75 & 68.25 & 47.25 & 69.25 & \textit{fix(24.79), docs(24.21), chore(18.00), feat(10.71), refactor(4.86), test(4.71), site(4.00)} \\
      playwright & 84K\ding{73} & \ding{52} & 58.00 & 46.84 & 47.00 & 64.25 & 41.00 & 74.75 & \textit{chore(24.86), feat(19.79), fix(17.14), docs(14.00), test(9.64), browser(8.29), devops(5.57) }  \\
      nest & 75K\ding{73} & \ding{52} & 70.75 & 51.43 & 64.00 & 77.25 & 12.25 & 84.75 & \textit{chore(54.79), fix(26.57), unknown(5.64), feat(5.36), test(3.00), refactor(2.64), sample(1.36) }  \\
      gatsby & 56K\ding{73} & \ding{52} & 52.50 & 43.61 & 46.50 & 68.25 & 32.75 & 67.00 & \textit{chore(46.64), fix(32.29), feat(11.07), docs(7.57), test(0.86), refactor(0.86), perf(0.71) }  \\
      uni-app & 41K\ding{73} & \ding{52} & 57.50 & 44.00 & 42.25 & 58.50 & 30.50 & 74.25 & \textit{chore(33.14), fix(24.21), wip(18.50), feat(16.79), release(3.64), refactor(2.21), build(1.50) }  \\
      vuetify & 41K\ding{73} & \ding{52} & 71.00 & 58.50 & 53.75 & 84.50 & 51.75 & 82.75 & \textit{docs(36.43), fix(33.21), chore(13.14), feat(7.93), refactor(5.86), test(2.14), style(0.64) }  \\
      sentry & 43K\ding{73} & \ding{52} & 38.00 & 55.39 & 55.00 & 47.50 & 35.25 & 61.75 & \textit{feat(31.79), fix(28.29), ref(24.93), chore(8.71), test(2.14), build(1.07), perf(0.64) }  \\
      union & 74K\ding{73} & \ding{54} & 46.50 & 40.50 & 46.25 & 66.75 & 30.25 & 66.50 & \textit{feat(33.93), chore(30.64), fix(29.57), docs(2.64), refactor(2.50), ci(0.71) }  \\
      taro & 37K\ding{73} & \ding{54} & 54.50 & 56.71 & 60.25 & 74.25 & 48.00 & 74.00 & \textit{fix(46.50), feat(25.20), chore(16.50), docs(4.43), refactor(4.00), test(2.21), perf(0.43) }  \\
      appsmith & 39K\ding{73} & \ding{54} & 54.50 & 61.65 & 50.25 & 71.00 & 42.75 & 71.25 & \textit{fix(40.93), chore(31.64), feat(12.07), test(8.21), ci(5.79), docs(0.79), feature(0.57) }  \\
      RSSHub & 42K\ding{73} & \ding{54} & 68.00 & 74.94 & 77.75 & 88.50 & 45.75 & 87.75 & \textit{feat(31.36), chore(30.07), fix(25.21), docs(5.50), style(4.29), test(1.29), refactor(0.86)}  \\
      leetcode & 35K\ding{73} & \ding{54} & 93.50 & 57.48 & 83.75 & 95.25 & 80.25 & 95.25 & \textit{feat(86.71), chore(8.79), docs(2.29), fix(1.57), style(0.64)}  \\
      payload & 41K\ding{73} & \ding{54} & 54.50 & 55.35 & 60.50 & 75.50 & 43.75 & 73.50 & \textit{chore(45.21), fix(29.57), feat(13.36), docs(6.07), ci(2.36), test(2.07), templates(0.79)}  \\
      pnpm & 34K\ding{73} & \ding{52} & 57.25 & 68.75 & 64.50 & 76.50 & 32.50 & 74.25 & \textit{chore(32.21), fix(32.07), feat(12.21), refactor(8.14), test(5.71), docs(4.64), ci(2.29)}  \\
      influxdb & 31K\ding{73} & \ding{54} & 43.25 & 61.40 & 60.50 & 65.50 & 46.50 & 64.50 & \textit{fix(25.21), feat(25.21), chore(20.29), refactor(17.79), test(5.21), docs(3.07), build(1.93)}  \\
      angular-cli & 27K\ding{73} & \ding{52} & 56.25 & 42.86 & 44.50 & 75.50 & 58.50 & 79.00 & \textit{build(31.71),fix(22.50), refactor(11.43), feat(8.86), docs(6.71), chore(6.21), test(5.50)}  \\
      quasar & 27K\ding{73} & \ding{54} & 55.50 & 43.19 & 51.00 & 76.00 & 36.50 & 75.75 & \textit{feat(44.00),fix(29.36), chore(20.79), refactor(3.14), docs(1.50), perf(1.21)}  \\
      \hline
      obs-studio  & 70K\ding{73} & \ding{54} & 93.00 & 9.25 & 94.50 & 12.75 & 7.50 & 97.75 & \textit{ui(53.07), libobs(25.14), ci(11.64), cmake(5.07), docs(1.71), decklink(1.29), aja(1.14)}  \\
      cockroach & 32K\ding{73} & \ding{54} & 57.25 & 42.86 & 60.00 & 2.25 & 2.50 & 77.50 & \textit{sql(22.50), storage(8.14), roachtest(7.86), opt(6.00), ui(5.21), kvserver(4.79), server(3.14)}  \\
      \bottomrule
    \end{tabular}
  }
\end{table*}

One of the major challenges faced by developers during a commit is deciding on the required type.
To this end, we propose automatically inferring project-specific types and compare its applicability to
five other approaches mentioned below.
For this experiment, we select twenty popular GitHub repositories (based on \ding{73} count) from our experimental setup  with differing levels of adherence to CCS. For each project, we randomly select 2000 commits of which 1400 are used for training, 200 for validation, and
400 for evaluation.

\subsubsection*{Baselines}
We conduct our evaluation using the following configurations:
\begin{enumerate}
  \item A state-of-the-art retrieval-based commit generation model, NNGen~\citep{liu2018neural}. For this experiment, we reuse the code shared by the authors and train it on our dataset.
  \item A developer-focused LLM-based approach in a zero-shot configuration using GitHub Copilot~\citep{chen2021evaluating}. To simulate a development workflow for commit message generation with Copilot, we clone the repository, check out the parent commit of the target commit, and provide an explicit prompt to generate a conventional commit message from the code diff.
  \item A general-purpose LLM-based approach in a zero-shot configuration using Gemini 2.0 model.
  \item An approach that uses type definitions manually curated in prior work~\citep{zeng2024first}. In this case, the commit-intent representation $\mathbb{C}$ is augmented with manual type definitions. We refer to this configuration as $\text{manual}_{\text{def}}$.
  \item A global variant of \tn{}, where the $\text{LLM}_{type}$ model is fine-tuned on commits sampled from different projects. We refer to this configuration as $\text{global}_{\text{def}}$.
  \item \tn{} in its default configuration, i.e., with $\text{LLM}_{type}$ fine-tuned on commits from the same repository.
\end{enumerate}

Table~\ref{tab:model_performance_across_projects} illustrates the results of our approach. For each case, the target label type is obtained by masking the corresponding part in the commit.  Overall, it shows the accuracy (in percentage) in generating the correct type for six configurations across twenty projects (top-20 rows of the Table).

\paragraph{Comparison to manual approach.}
CCS does not provide explicit definitions for the types used in commit messages. As a result, developers often struggle to correctly
assign appropriate types.
Manually providing definitions for these types is a time-consuming and non-scalable approach, particularly for large or rapidly evolving projects.

To address this limitation, we compare \tn{} to an approach  that additionally
contains manually created type definitions ($\text{manual}_{\text{def}}$).
Across all projects, our results indicate that \tn{} either outperforms or achieves comparable accuracy to the manual variant.
These finding demonstrate the generalizability of \tn{} and the applicability in a development workflow.

To further assess the generalizability and robustness of \tn{}, we evaluate its performance on two projects in which the most frequent commit types are largely absent from the manually curated type definitions.
The
corresponding
results are presented in the final two rows of Table~\ref{tab:model_performance_across_projects}.  These projects are not included in the above twenty because although the commits follow the structure of a CCS, the developers do not strictly mention CCS-compliance in the documentation.
We find \tn{} significantly outperforms its manually augmented counterpart; for instance, in the Cockroach project, it achieves an improvement by a factor of 34. This performance gap underscores the  ability of our approach to infer and generalize commit types where human-defined type mappings are unavailable and further highlight the limitations of a manual approach.

\paragraph{Comparison to baselines.}
\tn{} significantly outperforms all baseline approaches.
For instance, in the case of the deno project, which does not provide CCS guidelines for developers, \tn{} outperforms NNGen, Copilot, and Gemini by factors of 1.6, 1.3, and 1.26, respectively. Additionally, \tn{} outperforms the model trained on commits sampled from different projects across all projects in our dataset. These results demonstrate that type semantics are primarily project-specific and highlight the limitations of general-purpose and global models in capturing such semantics.

The last column of Table~\ref{tab:model_performance_across_projects} reports the top seven types and their frequencies (in percentage) for each project.
It highlights the prevalence of custom types among developers. For example, types such as \textit{site}, \textit{devops}, \textit{browser}, \textit{sample}, and \textit{unknown} are not part of CCS and are custom-created by developers.

Additionally, among the twenty libraries used in this experiment, only ten provide explicit guidelines for developers (third column of Table~\ref{tab:model_performance_across_projects}). The absence of agreed-upon CCS guidelines creates inconsistencies and confusion among developers about which type to select during a commit. In some cases, even with explicit guidelines, developers fail to adhere to them. The types listed for the \textit{freeCodeCamp} project, which includes guidelines (\ding{52}), provide one such example. Among the listed types are two that are essentially equivalent, namely \texttt{feat} and \texttt{feature}.
According to the CCS specification, \texttt{feat} should be used to indicate the introduction of a new feature, which is aligned with the project's guidelines.
However, developers frequently miss both recommendations and use \texttt{feature}, leading to inconsistencies and further highlighting the need for~\tn{}.

\subsection{RQ2: Effectiveness in Generating Description}

Apart from the type, another mandatory component of a commit conforming to CCS is the description.
We evaluate the capability of~\tn{} to generate the required description and compare it with four other approaches: (1) two general-purpose LLMs, Gemini 2.0 and GPT-4o, in a few-shot setting; for each model, we provide two example commits along with a manually crafted description as prompts to guide description generation for the target (queried) commit; (2) one state-of-the-art commit message generation approach, NNGen; and (3) one code-focused LLM, Copilot.

For this experiment, we sample a dataset of 2,000 CCS-compliant commits from our experimental setup, using 1,400 for training, 200 for validation, and 400 for evaluation. Table~\ref{tab:description_generation} reports the performance of~\tn{} in generating descriptions using the $\text{LLM}_{description}$ component. We find that a significantly smaller model fine-tuned for description generation outperforms both much larger general-purpose models, as well as the code-focused and state-of-the-art baselines.

\begin{table}[!ht]
  \centering
  \caption{Description generation performance against against baselines.}
  \label{tab:description_generation}
  \begin{tabular}{l|ccccc}
    \toprule
    Metric  & GPT-4o & Gemini & NNGen& Copilot &
    \makecell{Com-\\Beautify}\\
    \midrule
    Cosine Similarity & 0.2348 & 0.2964 & 0.0761 & 0.1980 &\textbf{0.3612} \\
    BLEU-2            & 0.1689 & 0.2476 & 0.0606& 0.0647&\textbf{0.3495} \\
    BLEU-3            & 0.1126 & 0.2030 & 0.0372 & 0.0360&\textbf{0.3105} \\
    ROUGE-L           & 0.3716 & 0.3716 & 0.1421 &0.1758&\textbf{0.4715} \\
    BERTScore F1      & 0.2637 & 0.2435 & -0.0738 &0.0247&\textbf{0.3958} \\
    \bottomrule
  \end{tabular}
\end{table}

\subsection{RQ3: Case-studies on the Benefits of Conventional Commits}\label{subsec:case_studies_ccs}
The conventional commit specification (CCS) proposes the  benefits~\footnote{\url{https://www.conventionalcommits.org/en/v1.0.0/\#why-use-conventional-commits}} below to developers for following the specification:
\begin{enumerate}
  \item Automatically generating CHANGELOGs.
  \item Automatically determining a semantic version bump.
  \item Triggering build and publish processes.
  \item Making it easier for people to contribute to your projects, by allowing them to explore a more structured commit history.
  \item Communicating the nature of changes to teammates, the public, and other stakeholders.
\end{enumerate}

To check if the proposed benefits are actually reaped by the developers, we perform an analysis of the varied usages of conventional commits. Of the benefits presented above, the last two are meant for other developers, while the first three are meant for machine consumption i.e., after a commit, it is read by a software tool and a range of predefined actions are performed.
GitHub Actions is a popular Continuous Integration and Continuous Deployment (CI/CD) platform that allows developers to automate software workflows based on different events. A typical GitHub Action is triggered when code changes containing one or more commits are pushed, starting automated workflows that perform tasks such as testing or  code quality checks. We present a case-study of GitHub Actions used by developers for conventional commits, while focusing only on the benefits proposed by the specification.

\subsubsection{Automatically generating CHANGELOGs}
\begin{figure}[!t]
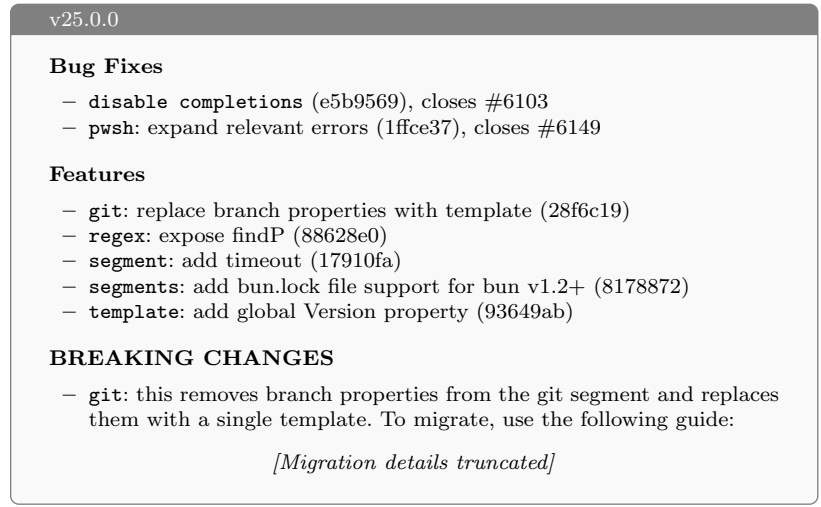

  \centering
  \begin{tcolorbox}[colback=gray!5!white, colframe=black!50, title=v25.0.0, width=0.9\linewidth, boxrule=0.5pt,  fontupper=\small]
    \textbf{Bug Fixes}
    \begin{itemize}
      \item \texttt{disable completions} (\href{https://github.com/JanDeDobbeleer/oh-my-posh/commit/e5b9569}{e5b9569}), closes \#\href{https://github.com/JanDeDobbeleer/oh-my-posh/issues/6103}{6103}
      \item \texttt{pwsh}: expand relevant errors (\href{https://github.com/JanDeDobbeleer/oh-my-posh/commit/1ffce37}{1ffce37}), closes \#\href{https://github.com/JanDeDobbeleer/oh-my-posh/issues/6149}{6149}
    \end{itemize}

    \vspace{0.5em}
    \textbf{Features}
    \begin{itemize}
      \item \texttt{git}: replace branch properties with template (\href{https://github.com/JanDeDobbeleer/oh-my-posh/commit/28f6c19}{28f6c19})
      \item \texttt{regex}: expose findP (\href{https://github.com/JanDeDobbeleer/oh-my-posh/commit/88628e0}{88628e0})
      \item \texttt{segment}: add timeout (\href{https://github.com/JanDeDobbeleer/oh-my-posh/commit/17910fa}{17910fa})
      \item \texttt{segments}: add bun.lock file support for bun v1.2+ (\href{https://github.com/JanDeDobbeleer/oh-my-posh/commit/8178872}{8178872})
      \item \texttt{template}: add global Version property (\href{https://github.com/JanDeDobbeleer/oh-my-posh/commit/93649ab}{93649ab})
    \end{itemize}

    \vspace{0.5em}
    \textbf{BREAKING CHANGES}
    \begin{itemize}
      \item \texttt{git}: this removes branch properties from the git segment and replaces them with a single template. To migrate, use the following guide:
    \end{itemize}
    \vspace{0.3em}
    {\centering \textit{[Migration details truncated]} \par}
    \vspace{0.3em}
  \end{tcolorbox}
  \caption{An example showing automatically generated changelog for Oh My Posh v25.0.0.}
  \label{fig:changelog_v2500}
\end{figure}

Figure \ref{fig:changelog_v2500} presents an example of an automatically generated changelog from the \href{https://github.com/JanDeDobbeleer/oh-my-posh}{oh-my-posh} (19.1K \ding{73}) GitHub repository. This changelog was produced using the \href{https://github.com/marketplace/actions/conventional-changelog-action}{Conventional Changelog GitHub Action}, a tool designed to streamline changelog generation from conventional commits or other  \href{https://github.com/conventional-changelog/conventional-changelog/tree/master/packages}{presets}. For  the \textit{conventionalcommits} preset, this action processes commit messages, grouping them by CCS types such as \texttt{feat} (new features), \texttt{fix} (bug fixes), and \texttt{revert} (reverted changes). For each identified group, it extracts and lists the one-line descriptions from the relevant commit messages. Furthermore, the action specifically identifies the \texttt{BREAKING CHANGE} keyword within commit messages and aggregates these into the BREAKING CHANGE section of the changelog.

Table \ref{tab:top_actions} lists the top three GitHub Actions by the number of repositories using them.
The \textit{semantic-pull-request} action ensures pull requests matches Conventional Commit Spec whereas \textit{Conventional Changelog action} and \textit{Release Please Action} both perform automatic changelog generation with Release Please Action additionally also perform semantic version bump.
The popularity of the  GitHub Action for automated changelog  generations, in particular from CCS commits, confirms the practical demand and benefits developers gain from using the CCS for this purpose.

\begin{table}[ht]
  \centering
  \caption{Top 3 Github Actions}
  \label{tab:top_actions}
  \begin{tabular}{|l|c|}
    \hline
    Action & \makecell{Number of Dependent\\ Repos} \\
    \hline
    semantic-pull-request & 46,849 \\
    \hline
    Conventional Changelog action & 14,422 \\
    \hline
    Release Please Action & 13,789 \\ %
    \hline
  \end{tabular}
\end{table}

\paragraph{Breaking change detection}
We notice that one of major components in automatic changelog generation is to determine if a commit introduces a breaking change.  The CCS defines a clear mechanism for communicating breaking changes in version control systems. Specifically, if a commit introduces a breaking API change, it can either append an exclamation mark $!$ after the type or scope (e.g., type$!$ or type(scope)$!$) or include a BREAKING CHANGE: tag followed by optional explanation in the footer.

We perform an experiment to check if the \textit{commit-intent-representation} of \tn{} is useful in detecting commits as breaking change. For this, we fine-tune a local model using \textit{commit-intent-representation} selected from the top-20 projects of Table~\ref{tab:model_performance_across_projects}. To reflect real-world commit distributions, where breaking changes are relatively rare, we construct dataset using 1:3 ratio of breaking to non-breaking commits. For evaluation, we remove any hints (e.g., remove $!$ from type$!$) present in a commit message about the breaking change and compare the performance of a general-purpose model (Gemini 2.0) and our fine-tuned model in identifying if a commit is a breaking change. We find that both  models perform relatively well where Gemini achieves an accuracy of 83.25\% while our model achieves 88\% (despite its relatively smaller size). The result demonstrates the usefulness of commit-intent-representation and also the applicability of a specialized model in automatically identifying a breaking change that may aid developers.

\subsubsection{Automatically determining a semantic version bump}
Beyond changelog generation, conventional commits also enable automated semantic version bump through predefined rules that map commit types to version increments. The semantic versioning specification defines three version components (MAJOR.MINOR.PATCH), and conventional commits can be used to determine which component should be incremented based on these rules:
\begin{itemize}
  \item PATCH (x.x.X): \texttt{fix:} commits - backwards compatible bug fixes
  \item MINOR (x.X.x): \texttt{feat:} commits - new backwards compatible features
  \item MAJOR (X.x.x): Any commit with \texttt{BREAKING CHANGE:} in footer or \texttt{!} after type/scope - incompatible API changes
\end{itemize}
Figure~\ref{fig:version_bump_example} shows two release notes from the \href{https://github.com/JanDeDobbeleer/oh-my-posh}{oh-my-posh} repository. A breaking change was introduced after version 25.23.3, so the major version was incremented in the next release.
The Github actions
in Table \ref{tab:top_actions} can also be used to bump the semantic version.

\subsubsection{Impact on Commits}
Commits by developers that bundle multiple unrelated changes (fixing a bug and also making changes to the documentation) create a \textit{tangled}  commit~\citep{dias2015untangling,herzig2011untangling}. Tangled commits are discouraged~\citep{kondo2024empirical} because such commits make code reviews and reverting changes difficult, and make code merging more complex. Instead, atomic commit that makes one logical change to the codebase is encouraged.

CCS mandates using a single word to indicate the type and a brief sentence to summarize the change. To this end, we  measure the implications of enforcing such constraints on the commit size. Specifically, we examine if developers adhering to CCS tend to make commits involving smaller code changes.

In this experiment, we randomly select a sample of 100,000 commits stratified over repositories that adhere to CCS, and 100,000  that do not, drawn from our experimental dataset.
The analysis reveals that the average number of lines per commit is 466 for CCS-compliant commits and 679 for non-compliant ones. The results show that the commits that follow  CCS are on average  30\% smaller in size. These findings suggest that developers who adhere to CCS are more inclined to make smaller, more incremental code changes, further reinforcing the potential benefits of CCS.

\begin{figure}[t]
  \centering
  \begin{tcolorbox}[
      title=\textbf{v25.23.3 - Release Notes},
      colback=yellow!2!white,
      colframe=black!50,
      coltitle=black,
      fonttitle=\bfseries,
      fontupper=\small,
      enhanced,
      boxrule=0.5pt,
      sharp corners
    ]
    \textbf{Bug Fixes}
    \begin{itemize}
      \item \textbf{schema:} remove format from color\_string \href{https://github.com/JanDeDobbeleer/oh-my-posh/commit/1e9da59b4b5ed314461df66650170a37aeecfd65}{(\textcolor{blue}{1e9da59})}, closes \href{https://github.com/JanDeDobbeleer/oh-my-posh/issues/6457}{\textcolor{blue}{\#6457}}
      \item \textbf{upgrade:} display from version - to version \href{https://github.com/JanDeDobbeleer/oh-my-posh/commit/b919375dc6c7e7ea14fd2806f35d652e9659817c}{(\textcolor{blue}{b919375})}, closes \href{https://github.com/JanDeDobbeleer/oh-my-posh/issues/6439}{\textcolor{blue}{\#6439}}
    \end{itemize}
  \end{tcolorbox}
  \begin{tcolorbox}[
      title=\textbf{v26.0.0 - Release Notes},
      colback=green!5!white, %
      colframe=black!50,
      coltitle=black,
      fonttitle=\bfseries,
      fontupper=\small,
      enhanced,
      boxrule=0.5pt,
      sharp corners
    ]
    \textbf{Bug Fixes}
    \begin{itemize}
      \item \textbf{scm:} sort mapped\_branches alphabetically \href{https://github.com/JanDeDobbeleer/oh-my-posh/commit/e632705586a26f27adeaa826c9ef2c480250f287}{(\textcolor{blue}{e632705})}, closes \href{https://github.com/JanDeDobbeleer/oh-my-posh/issues/6469}{\textcolor{blue}{\#6469}}
    \end{itemize}
    \textbf{Features}
    \begin{itemize}
      \item \textbf{aws:} add shortened .RegionAlias \href{https://github.com/JanDeDobbeleer/oh-my-posh/issues/6469}{(\textcolor{blue}{0cd4f42})}
      \item[\vdots] \textit{Additional features omitted for space}
    \end{itemize}
    \textbf{BREAKING CHANGES}
    \begin{itemize}
      \item \textbf{init:} this requires Nushell users to remove \texttt{source \textasciitilde/.oh-my-posh.nu} from their \texttt{\$nu.config-path} file.
    \end{itemize}
  \end{tcolorbox}
  \caption{An example commit showing major version bump when a BREAKING CHANGE is released.}
  \label{fig:version_bump_example}
\end{figure}

\subsection{RQ4: Utility of Code Summarization}
As an ablation study, we evaluate if performing an additional step to summarize the code differences affects the performance of~\tn{}.
To this end, we compare~\tn{} to two other approaches. The first approach is a prior work by \citet{zeng2024first} (referred to as Non-summarized), that uses an LLM fine-tuned  directly on unprocessed  commits to classify CCS types;
The second approach uses   zero-shot inference using the GPT-4o model. In the latter case, we use prompting techniques to generate types from the GPT-4o without any additional learning involved. Since the approach proposed in prior work~\citep{zeng2024first} also incorporates manual definitions of type, for this experiment, we provide the manual definitions as part of the input for all of the approaches to ensure fair comparison.

The dataset used for this experiment contains 2000 manually curated conventional commits from  existing work~\citep{zeng2024first} of varying types, of which 1400 are used for training, 200 for validation, and 400 for testing.
Since GPT-4o does not involve any
training step, we use the 400 commits in the test dataset to report the results.

Table~\ref{tab:generating_type_performance} summarizes the results of this experiment. The results show that the
$\text{LLM}_{summary}$ component of~\tn{} is indeed useful,  outperforming both alternative approaches. We conclude that providing a natural language summary, instead of using a diff which can be arbitrarily large,  improves the overall performance in generating a CCS type.
\begin{table}[ht]
  \centering
  \caption{Performance of \tn{} on generating type evaluated on a manual dataset curated by a previous work\cite{zeng2024first}.}
  \label{tab:generating_type_performance}
  \begin{tabular}{l|cccccc}
    \toprule
    \textbf{Metric} & \textbf{GPT-4o} & \textbf{Non-summarized} & \textbf{\tn{}} \\
    \midrule
    Accuracy         & 73.75\% & 76.50\% &\textbf{78.50\%} \\
    Macro Precision  & 75.17\% & 77.95\% &\textbf{78.07\%} \\
    Macro Recall     & 73.75\% & 76.50\% &\textbf{78.50\%} \\
    Macro F1 Score   & 73.29\% & 76.41\% &\textbf{78.21\%} \\
    \bottomrule
  \end{tabular}
\end{table}

\section{Practical Applicability of \tn{}}
Based on the growing trend of CCS adoption, its usefulness and the encouraging evaluation results of~\tn{}, we envision several practical use-cases for our approach:
\begin{itemize}
  \item Generation of CCS commits within code editors, from developer's initial draft messages.
  \item Automated semantic versioning and CHANGELOG generation in repositories with mixed commits.
  \item Unified classification and processing of legacy commits for mining software repositories in software engineering research and statistical analysis.
\end{itemize}

One of the primary benefits of \tn{} is its automated approach in generating commits that conform to CCS. Deploying \tn{} for a project requires an one time effort of training using a sample of past commits. Once trained,
we envision that \tn{} will be deployed as a service i.e., for a given project, a dedicated workstation will handle requests from the developers.
Additionally, we demonstrate that even using significantly smaller models, \tn{} outperforms the current state-of-the-art LLMs. This results in a reduced overall hardware requirement, making the system more practical and efficient to deploy.

\section{Threats to Validity}\label{sec:threats_to_validity}
CCS is a relatively recent proposal and its adoption is evolving rapidly. Consequently, the trends observed in our current analysis may not remain valid in the future. To address this threat, we take a long-term view of the trend starting from 2017 through 2025. 
Our project-specific evaluation is based on twenty open-source projects, which may not constitute a representative sample. We address this concern by choosing popular and diverse projects each with at least 30K GitHub stars. To identify if a commit complies with CCS, we rely on the structure as specified by CCS. However, it is possible for commits to follow the prescribed format without fully adhering to the intent of the specification. To reduce the implications of this threat, we manually inspect a sample of commits and find them to be CCS compliant. We conduct our experiments utilizing specific versions of LLMs. LLMs are known to hallucinate results which can have an impact on our results. Additionally, any future updates to these models can adversely impact our findings. To address this threat, we preserve the findings of our experiments. LLMs are also known to be sensitive to prompts and as a result, the a major threat for our approach comes from crafting the prompts. 

\section{Related Work}\label{sec:related_work}

Our work on generating CCS-compliant commits contributes to an active research field of automated Commit Message Generation (CMG),
a functionality now widely integrated into code editors as part of AI code assistance.
CMG aims to assist developers by automatically producing informative commit messages, 
and incorporates various retrieval-based and learning-based techniques \citep{Zhang2024}.
Recently, large language models (LLMs) have been actively  applied to CMG.
For instance,
~\citet{Xue2024} studied the use of LLMs for CMG and proposed ERICommiter, a framework that curates and retrieves relevant historical commit examples to effectively guide LLMs via in-context learning to generate high-quality commit messages.
Similarly,
\citet{Zhang2024a} proposed REACT, a retrieval-based framework that finds exemplar diff-message pairs to guide LLMs during CMG.
The efficacy of in-context learning for LLM-based CMG was further empirically investigated by
~\citet{Wu2025}, who studied the impact of various prompts and demonstration settings across different LLMs and datasets.
Other research has focused on enabling LLMs to leverage richer  context beyond code diffs or simple exemplars.
~\citet{Li2024} introduced Omniscient Message Generator, which treats CMG as a knowledge-intensive reasoning task and employs an LLM to incorporate diverse software context (e.g., issue reports, pull requests) for generating more comprehensive commit messages.
\citet{Eliseeva2023} investigated history-aware commit message completion.
 This use of past commit history aligns with our intuition behind generating project-specific CCS  messages.
Compared to our work, these CMG approaches do not adhere to the structured CCS specification; however, the proposed techniques for richer contextual and historical understanding could potentially improve  generation of CCS descriptions.

Another relevant line of research targets commit classification, which is relevant to the task of   deriving the CCS-compliant commit type.
Numerous studies have focused on classifying commits using techniques ranging from word frequency analysis \citep{mockus2000identifying,hindle2009automatic}, feature extraction with NLP models \citep{levin2017boosting,fluri2007change,ghadhab2021augmenting,feng2020codebert,lee2021co}, to generative models \citep{Sazid2024,Tong2025}. 
A recent work by
\citet{herivcko2023commit} reviews various commit classification techniques for software maintenance tasks.

The first systematic research focused specifically on Conventional Commits Specification (CCS)
is conducted by ~\citet{zeng2024first}. The work studies CCS adoption,
identifies challenges developers face when selecting CCS types,
and proposes an LLM-based approach to classify commits into ten predefined
CCS types. In contrast, our paper focuses on the generation of
CCS-compliant commit types and descriptions.
Furthermore, while \citet{zeng2024first}  relies on a manually curated list
of CCS types,
our study demonstrates the importance of project-specific conventions,
showing that models fine-tuned on project-specific data outperform generalized models. Additionally, we include a considerably larger analysis of CCS adoption,
based on over 20 million commits, and provide case studies that highlight real-world benefits of using CCS.

\section{Conclusion}\label{sec:conclusion}

Version control systems are central to modern software development, enabling collaboration and tracking code changes. The recently introduced Conventional Commit Specification (CCS) enforces a structured commit. The aim of~\tn{} is to automatically synthesize a complete structured commit given a code change and a natural language description. We address the fundamental challenge faced by developers in classifying commits leveraging two key insights. First, instead of using code differences directly, we propose to use a summary of the change. Second, instead of relying on manual annotations of commits, our approach automatically learns from historical commits. 
Our evaluation on real-world projects shows improvements in predicting the correct CCS type over an approach that uses manual annotation of definitions. Additionally, for generating the description, our approach  outperforms two larger general-purpose models.
We envision~\tn{} being used by developers to  alleviate some of the challenges of correctly assigning a class to a commit.
Beyond generating the type and description of a commit, future efforts could focus on generating the body as well. 
Additionally, extending~\tn{} to generate the scope alongside the type for a CCS commit represents another potential direction
worth investigating.

\section{Artifact Availability}
We make the code and associated data of this work available at:\\ \href{https://figshare.com/s/c36dfd3195676aeb9cef}{https://figshare.com/s/c36dfd3195676aeb9cef}
 
\bibliographystyle{spbasic}      %
\bibliography{ref}
\end{document}